\documentclass[preprint,showpacs,groupedaddress,amsmath,amssymb]{revtex4}
\usepackage{graphicx}
\usepackage{dcolumn}
\usepackage{multirow}
\usepackage{amsfonts}

\newcommand{\beq}{\begin{equation}}
\newcommand{\eeq}{\end{equation}}
\newcommand{\bea}{\begin{eqnarray}}
\newcommand{\eea}{\end{eqnarray}}

\newcommand{\vepsi}{\epsilon}

\newcommand{\bhat}{\hat{b}}
\newcommand{\hhat}{\hat{h}}

\newcommand{\zhat}{\hat{z}}
\newcommand{\thetahat}{\hat{\theta}}

\newcommand{\wt}[1]{\widetilde{#1}}
\newcommand{\mbf}[1]{\mathbf{#1}}

\newcommand{\del}{\nabla}
\newcommand{\divr}{\nabla \cdot}
\newcommand{\curl}{\nabla \times}
\newcommand{\DDt}[1]{\dfrac{D\, {#1}}{D t}}
\newcommand{\ddt}[1]{\dfrac{\partial\, {#1}}{\partial t}}

\newcommand{\dext}{\mathrm{d}}
\newcommand{\ptild}{\widetilde{p}}

\newcommand{\MoH}{\dfrac{M}{H}}

\makeatletter
\def\Tbar{\mathchoice
   {\TTbar\displaystyle\textstyle{-}}%
   {\TTbar\textstyle\scriptstyle{-}}%
   {\TTbar\scriptstyle\scriptscriptstyle{-}}%
   {\TTbar\scriptscriptstyle\scriptscriptstyle{-}}%
   \!T}
\def\TTbar#1#2#3{{\setbox0=\hbox{$#1{#2#3}{\mathrm{T}}$}
     \raise2\p@\vbox{\hbox{$#2#3$}}\kern-.35\wd0}}
\makeatother

\begin{document}

\title{On the application of Maxwell's theory to many-body systems, or why the resistive magnetohydrodynamic equations are not closed}


\author{Robert W. Johnson}
\email[]{rob.johnson@gatech.edu}
\affiliation{Atlanta, GA 30238, USA}

\date{November 25, 2008.}

\begin{abstract}
The resistive magnetohydrodynamic (MHD) equations as usually defined in the quasineutral approximation refer to a system of 14 scalar equations in 14 scalar variables, hence are determined to be complete and soluble.  These equations are a combination of Navier-Stokes and a subset of Maxwell's.  However, one of the vector equations is actually an identity when viewed from the potential formulation of electrodynamics, hence does not determine any degrees of freedom.  Only by reinstating Gauss's law does the system of equations become closed, allowing for the determination of both the current and mass flow velocity from the equations of motion.  Results of a typical analysis of the proposed electromagnetic hydrodynamic model including the magnetization force are presented.
\end{abstract}

\pacs{47.65.-d, 52.30.-q, 52.55.-s} 

\maketitle


The resistive magnetohydrodynamic (MHD) equations as usually defined in the quasineutral approximation refer to a system of 14 scalar equations in 14 scalar variables, hence are determined to be complete and soluble.  These equations are a combination of Navier-Stokes and a subset of Maxwell's.  However, one of the vector equations is actually an identity when viewed from the potential formulation of electrodynamics, hence does not determine any degrees of freedom.  Only by reinstating Gauss's law does the system of equations become closed, allowing for the determination of both the current and mass flow velocity from the equations of motion.  Results of a typical analysis of the proposed electromagnetic hydrodynamic model including the magnetization force are presented.

Many authors~\cite{chen-84,dendybook-93,dinkbook-05,staceybook05,kivel95,spsbds03,mbk04,goldruth95} define the low frequency resistive MHD equations as the zeroth and first order moments of the Vlasov equation with adiabatic closure in conjunction with the two curl equations among Maxwell's.  For the neutral fluid, the sum and difference of the ion and electron equations of motion give the net force balance equation and the generalized Ohm's law.  Using $D/Dt \equiv \partial / \partial t + \mbf{V}_f \cdot \del$, we write the usual equations: \bea \label{eqn:1}
\ddt{\rho_m} + \divr (\rho_m \mbf{V}_f) = 0 \;, & & \DDt{p/ \rho_m^\gamma} = 0 \;, \\ \label{eqn:2}
\rho_m \DDt{\mbf{V}_f} = \mbf{J} \times \mbf{B} - \del p \;, & & \eta \mbf{J} = \mbf{E} + \mbf{V}_f \times \mbf{B}\;, \\ \label{eqn:3}
\curl \mbf{E} = - \ddt{\mbf{B}} \;, & & \curl \mbf{B} = \mu_0 \mbf{J}\;,
\eea where $\eta$ is the resistivity and $\gamma$ is the appropriate index for the case under consideration, and the degrees of freedom are pressure $p$, mass density $\rho_m$, flow velocity $\mbf{V}_f$, current $\mbf{J}$, and electromagnetic fields $\mbf{E}$ and $\mbf{B}$, giving a naive counting of 14 scalar equations for 14 scalar variables.  However, while for decades~\cite{brag-1965,roseclark} the argument has been made that Gauss's law may be neglected with impunity, no one within the plasma physics community has denied the applicability of the potential formulation of electrodynamics~\cite{maxwell-1864}.  The unnamed of Maxwell's equations (often called the ``no-monopole'' equation, $\divr \mbf{B} = 0$) is brought into play during the determination of plasma equilibrium, via solution of the Grad-Shafranov equation~\cite{grad-58,shafranov-66} in toroidal geometry or otherwise, which by the naive counting of above would introduce an additional scalar equation, thus over-determining the system, yet is commonly known simply to allow for the expression of the magnetic field in terms of the vector potential, $\mbf{B} = \curl \mbf{A}$.  The reason doing so is valid is because vector identities by mathematical definition do not determine any degrees of freedom; they reduce them.  Inserting that expression into Faraday's law~\cite{griffiths-89}, we recover $\curl (\mbf{E} + \partial \mbf{A} / \partial t) = 0$, whence $\mbf{E} = -\del \Phi - \partial \mbf{A} / \partial t$, which clearly displays the division of the electric field into static and dynamic components and reduces three of our naive degrees of freedom down to one for which we have no equation.  Unless one wishes to invent new physics, the resolution is clear---the reinstatement of Gauss's law, $\divr \mbf{E} = - \del^2 \Phi - \partial (\divr \mbf{A})/\partial t = \rho_e/\vepsi_0$ which vanishes for a neutral fluid, is required to close the system of equations, bringing the number of scalar equations and degrees of freedom into agreement with the number $14-3+1=14-2=12$.  We remark that Faraday's law is no less an identity than the no-monopole equation as both are given by the general theory of vector fields.  Gauge invariance plays a special role in the local conservation of charge, best expressed in manifestly Lorentz covariant notation.

From a particle physicist's field-theoretic point of view~\cite{halzenmartin,davis70,ramond-FTM90,ryder-qft,mandlshaw}, the Maxwell field tensor $F^{\mu \nu} \equiv \partial^\mu A^\nu - \partial^\nu A^\mu$ in media is known to have only 3 physical degrees of freedom embodied by the four-potential $A^\mu \equiv (\Phi/c,\mbf{A})$ subject to the gauge condition, not 3 for each of the electric and magnetic fields, which couple to sources given by the conserved four-current $J^\mu \equiv (c \rho_e,\mbf{J})$ through the inhomogeneous Maxwell equations $\partial_\mu F^{\mu \nu} = (\partial_\mu \partial^\mu) A^\nu - \partial^\nu (\partial_\mu A^\mu)  = \mu_0 J^\nu$, which are explicitly Lorentz covariant and also gauge invariant, and the homogeneous Maxwell equations, given by the divergence of the dual tensor $\wt{F}^{\mu \nu} \equiv \vepsi^{\mu \nu \alpha \beta} F_{\alpha \beta}/2$, where $\vepsi^{\mu \nu \alpha \beta}$ is the permutation tensor, as $\partial_\mu \wt{F}^{\mu \nu} = 0$, are satisfied identically when written in terms of the electromagnetic potential, hence do not determine any degrees of freedom.  Antisymmetry in $F^{\mu \nu}$ immediately implies conservation of the current, $\partial_\nu \partial_\mu F^{\mu \nu} = \mu_0 \partial_\nu J^\nu = 0$, thus it carries only 3 degrees of freedom also.  One may recast the Maxwell equations into a component-free form through the use of differential geometry~\cite{ryder-qft}, where ``the existence of integrals implies a duality between forms and chains'' which may be exploited.  In natural units $\mu_0 \equiv \vepsi_0 \equiv c \equiv 1$ and using the exterior derivative $\dext$, the Hodge dual $\;^*$, the connection 1-form $A \equiv A_\mu dx^\mu$, the curvature 2-form $F \equiv (-F_{\mu \nu}/2) dx^\mu \wedge dx^\nu$, and the current 3-form $J \equiv (J_x dy \wedge dz + J_y dz \wedge dx + J_z dx \wedge dy) \wedge dt - \rho_e dx \wedge dy \wedge dz$ which satisfies the continuity equation $\dext\, J = 0$, one writes the field equation as $\dext\, ^*\!F = J$ and the Bianchi identity, which is a statement on the structure of the manifold, as $\dext\, F = 0$, whence $F = \dext\, A$, and we remark that gauge invariance, through Noether's theorem, implies conservation of the covariant current.  What all this shows is that the natural, physical division of the Maxwell equations is not into the divergence and curl equations but rather into the homogeneous and inhomogeneous equations, whereby the Bianchi identity carries the structure for the potential formulation and the field equation carries the dynamics obtained from the action.

The implication for plasma physics is clear: the quasineutral approximation does worse than just neglect an effect, as it introduces inconsistency into the equations when the components of the electrostatic field are treated in isolation~\cite{rwj-08cpp02}.  Arguing that Maxwell's divergence equations are initial conditions for the curl equations is incorrect in media, for while in vacuum such statement leads to the propagation of electromagnetic radiation with two physical states of polarization, the source terms spoil such interpretation, and the divergence of the Maxwell-Ampere equation only recovers the equation for local charge conservation, which must be respected independently of the conservation of mass addresed by the zeroth moment of the Vlasov equation, when Gauss's law retains its intended form.  Note that authors not including the no-monopole equation explicitly within the system do not make the argument of having 14 equations and degrees of freedom, as that equation represents an additional member.  Claiming that in general the sources may be uniquely determined from expressions for the fields is inappropriate, for while suitable boundary conditions must be supplied, the differential operators hence the boundary conditions are applied to the fields, not the sources.  The reason for the expression ``Maxwell-Lorentz electrodynamics'' is because Maxwell's theory tells one how the fields react to the sources, and the Lorentz force through the equations of motion tells the sources how to react to the fields; trying to go the other way around the loop is not well defined, as the physics is contained within the action from which both the field and source equations of motion may be obtained.  

Let us examine in detail where difficulties are encountered by the neoclassical approach, a term we use to encompass all non-classical approaches to the fluid description of ionized particles regardless of geometry---such discussion~\cite{npg-14-49-2007} invariably engenders a hostile response~\cite{npg-14-543-2007,npg-14-545-2007} from its adherents yet is necessary if one is to consider the application of electrodynamic field theory in tensor notation to the many-body system commonly called a plasma.  The scalar degrees of freedom $\rho_m$ and $p$ may be associated with the scalar equations for mass and energy conservation, Equations~(\ref{eqn:1}), as no other quantities appear in those equations for the case of vanishing flow velocity; the presence of a flow velocity $\mbf{V}_f$ couples those equations to the rest of the system to be solved simultaneously.  Note that the previous argument of the second paragraph tacitly assumed that the equations of motion in the form of the generalized Ohm's law and the convective force balance, Equations~(\ref{eqn:2}), were associated with the degrees of freedom $\{\mbf{V}_f,\mbf{J}\}$; whereas here, without Gauss's law, one must determine the electric field from an equation of motion, usually the generalized Ohm's law (however the ion~\cite{solomonetal-pop-2006} and electron~\cite{frc-pop-2006} equations of motion are also used), giving the solution $\mbf{E}_{neo} = \eta \mbf{J} - \mbf{V}_f \times \mbf{B}$.    Faraday's law in conjunction with the no-monopole equation then relates the electric field to the potentials $-\mbf{E}_{neo} = \partial \mbf{A}/\partial t + \del \Phi$, where without Poisson's equation or its gauge invariant generalization the relation between the potentials and the space charge density $\rho_e$ remains unspecified (in essence, Faraday's law here {\it determines} a potential $\Phi$ which is not an independent degree of freedom), and its divergence gives in various gauges \bea
\divr \left( \mbf{V}_f \times \mbf{B} - \eta \mbf{J} \right) & = & \ddt{} \divr \mbf{A} + \del^2 \Phi \;, \\
\mathrm{Coloumb}\; (\divr \mbf{A} = 0)\; & = & \del^2 \Phi \;, \\
\mathrm{Lorenz}\; (\divr \mbf{A} = -\mu_0 \vepsi_0 \ddt{} \Phi)\; & = & \Box^2 \Phi \;, \\
\mathrm{Weyl}\; (\Phi = 0)\; & = & \ddt{} \divr \mbf{A} \;,
\eea where the LHS is explicitly gauge invariant whereas the form and interpretation of the RHS is dependent upon one's choice of gauge.  The issue of gauge invariance is a red herring in the discussion, for while true physics must be equally described in any and all gauges, the crucial error in the neoclassical approach is its use of an equation of motion to determine the electric field, which does not respect Lorentz covariance.  (Note that modern power generators and electric motors certainly are not moving materially at relativistic speeds yet make full and practical use of the covariant transformation properties of the field tensor through Faraday's law of induction.)  Returning to the expression for $\mbf{E}_{neo}$, let us now examine its transformation properties under a change of reference frame.  Let $S$ be the frame of the neoclassical observer, and let $S'$ be the frame moving with velocity $\mbf{V}_f$ with respect to $S$.  Without loss of generality, the flow velocity in $S$ is taken along the $x$-axis, thus $\mbf{V}_f= (V_f, 0, 0) \neq 0$ gives $\mbf{E}_{neo}=(\eta J_x, \eta J_y + V_f B_z, \eta J_z - V_f B_y)$, using Einstein's velocity addition rule~\cite{einstein-1905a} gives $\mbf{V}_f'=0$, and for $\gamma \equiv 1/\sqrt{1-V_f^2/c^2}$ the transformation for proper velocity applies to $\mbf{J}$, the spatial part of the four-current $J^\mu_{neo} = (0,\mbf{J})$, giving \beq
\mbf{E}_{neo}' = \eta' \mbf{J}' = \left[ \begin{array}{c} \eta' \gamma J_x \\ \eta' J_y \\ \eta' J_z \end{array} \right] \neq \left[ \begin{array}{c} \eta J_x \\ \gamma \eta J_y \\ \gamma \eta J_z \end{array} \right] = \mbf{E}' \;,
\eeq where $\mbf{E}' = [E_x, \gamma(E_y - V_f B_z), \gamma(E_z + V_f B_y)]$ is the transformation law for the physical electric field.  Equality could hold only if $\eta' = \eta / \gamma = \eta \gamma$ implying $\gamma = 1$, which holds only when $V_f = 0$, thus only in the neoclassical frame of reference but also implying a vanishing flow velocity, contradicting the initial assumption $V_f \neq 0$.  The expression for $\mbf{E}_{neo}$ has inherited the nature of a velocity vector from its neoclassical determination hence cannot possibly represent a true electric field, which {\it does not} transform as the spatial part of a four-vector~\cite{maxwell-1864,einstein-1905a,griffiths-89}.  Furthermore, as ultimately $\mbf{B}(\mbf{J})$ may be determined from Ampere's law or the equivalent Biot-Savart law (for steady currents only as all the terms with $\mbf{E}$ need be present for Maxwell's theory to respect local charge conservation), the neoclassical electric field depends explicitly on the two vectorial quantities of current and mass flow, $\mbf{E}_{neo}(\mbf{J},\mbf{V}_f)$.  In order to completely determine the system, both of those quantities must find solution; however, having already used one of our equations of motion in the guise of Ohm's law, we have left only one vector equation for the net conservation of momentum, $\rho_m D \mbf{V}_f / D t + \del p = \mbf{J} \times \mbf{B}$, which leaves one vector's worth of degrees of freedom without solution, leading to the use of a stationary equilibrium equation $\del p = \mbf{J} \times \mbf{B}$ in the analysis of non-stationary plasma experiments~\cite{frc-pop-2006,solomonetal-pop-2006}.  We note that the predictions of the neoclassical (NCLASS) model for the poloidal velocity found in a tokamak presented in Reference~\cite{solomonetal-pop-2006} explicitly fail to agree with the experimental measurements.  By reinstating the determination of the electrostatic field via Gauss's law, what returns is the generalized Ohm's law, an equation of motion which one may solve for the motion appearing in that equation, which in conjunction with the convective force balance equation fully determines the system.  Ultimately, the various arguments presented in support of the neglect of Gauss's law are superseded by the rigorous formalism of differential geometry, whereby casting the Maxwell equations into intrinsic, geometric form, $\dext\, ^* \dext\, A = J$, comprises very deep and powerful statements concerning what is known about our Universe.

Consequently, we here advocate the use of the electromagnetic hydrodynamic (EMHD) equations specified by the fluid equations for all species $s$, \bea \label{eqn:fluid1}
\ddt{n_s m_s} + \divr (n_s m_s \mbf{V_s}) = 0 \;, & & \left( \ddt{} + \mbf{V}_s \cdot \del \right) \dfrac{p_s}{(n_s m_s)^{\gamma}} = 0 \;, \\ \label{eqn:fluid2}
n_s m_s \left( \ddt{} + \mbf{V}_s \cdot \del \right) \mbf{V}_s + \del p_s &=& n_s e_s \left(\mbf{E} + \mbf{V}_s \times \mbf{B} \right) + \sum_k \mbf{F}_{s k}\;,
\eea associated with the degrees of freedom $\{n_s, \Tbar_s, \mbf{V}_s\}$ for $p_s = n_s \Tbar_s \equiv n_s k_B T_s$, where $k\neq s$ for the friction term $\mbf{F}_{s k} = - \mbf{F}_{k s}$ representing interspecies collisions and neglecting viscosity, polarization, and magnetization, in conjunction with the field equation $\partial_\mu F^{\mu \nu} = \mu_0 J^\nu$ associated with the degrees of freedom $A^\mu$.  Extension to incorporate the magnetic decomposition $\mbf{B}/\mu_0 = \mbf{H} + \mbf{M}$ into free and bound currents is straightforward, and a gyrotropic pressure tensor is forthcoming, noting that the fluid mechanics of Equations~(\ref{eqn:fluid1}) and (\ref{eqn:fluid2}) remains to be cast into the Lorentz covariant form manifest for the electromagnetic sector.  As a preliminary, we solve the Equations (\ref{eqn:fluid2}) for everyone's favorite hydrogenic plasma of fusion interest, fully ionized deuterium, in the neutral fluid limit $\rho_e \rightarrow 0$ such that $\divr \mbf{E} \rightarrow 0$, for the case of an infinite column at equilibrium $\partial / \partial t \rightarrow 0$ with prescribed coaxial applied electric and magnetic fields and an assumed pressure profile.  With the definitions for species $s \in \{e,i\}$ of particle density $n \equiv n_e + n_i = 2 n_0$, mass density $\rho_m \equiv \sum_s n_s m_s$, free momentum density $\rho_m \mbf{V}_f \equiv \sum_s n_s m_s \mbf{V}_s$, and free current density $\mbf{J}_f \equiv \sum_s n_s e_s \mbf{V}_s$, one finds \beq \label{eqn:VeVi}
\left[\begin{array}{c} \mbf{V}_e \\ \mbf{V}_i \end{array} \right] = \mbf{V}_f + \dfrac{\mbf{J}_f}{e \rho_m} \left[\begin{array}{c} - m_i \\ m_e \end{array}  \right] \;,
\eeq where $\mbf{V}_{e,i}$ are the species' fluid or guiding center velocities.  Gyromotion is so far incorporated through the inclusion of the drift momentum $\rho_m \mbf{V}_d \equiv \curl \mbf{L}_g$ as the curl of the angular momentum density carried by the cyclotron motion and the diamagnetic current $\mbf{J}_d \equiv \curl \mbf{M}$ as the curl of the magnetic moment per unit volume, giving total momentum density $\rho_m \mbf{V} \equiv \rho_m (\mbf{V}_f + \mbf{V}_d)$ and total current density $\mbf{J} \equiv \mbf{J}_f + \mbf{J}_d$, and through the magnetization force $\mbf{F}_M \equiv \mu_0 \del \mbf{M} \cdot \mbf{H} = - \mu_0 \del M H$ as $\mbf{M} = - M \hhat = - (M/H) \mbf{H}$, an important effect neglected in the usual analysis of plasma equilibrium despite its experimental applications in fusion~\cite{hayes-213}, magnetic fluids~\cite{rinaldi-2847,zahn-144,eldib-159}, biophysics~\cite{tzir-077103,rama-297,qi-132,ataka-ic0038,wang-877}, and materials science~\cite{maki-1132,maki-1096,maki-066106,ma-2944,asai-r1,takagi-842,kozuka-884,fangwei-024202,colli-58,ono-608,cantor-02,keil-07} and addressed for the case of a stationary equilibrium elsewhere\cite{rwj-ppcf03}.  For the gyro-momentum, we take $\mbf{L}_g \equiv \sum_s \mbf{L}_{g s} \equiv \sum_s n_s \vec{l}_s$ for $\vec{l}_s = m_s v_{\perp s}^2 \vec{\omega}_s / \omega_s^2 = 2 \Tbar_s \vec{\omega}_s / \omega_s^2$ where $\vec{\omega}_s \equiv - e_s \mbf{B}_s / m_s$ and $v_{\perp s} = \sqrt{2 W^\perp_s / m_s}$, and for the species dipole density $\mbf{M}_s \equiv n_s \vec{\mu}_s = -n_s (W^\perp_s / B_s) \hhat$, where the total field felt by a particle of species $s$ is $\mbf{B}_s / \mu_0 = \mbf{H}+\mbf{M}-\vec{\mu}_s = \mbf{H} + \mbf{M}_k + \alpha_s \mbf{M}_s \approx (H-M)\hhat$ for $\alpha_s \equiv (n_s - 1)/n_s \approx 1$ at sufficient plasma density and points along $\bhat_s \equiv \mbf{B_s}/B_s = \hhat$, giving a total magnetic dipole moment of $\mbf{M} \equiv \sum_s f_s \mbf{M}_s = 2 n_0 (f_e \vec{\mu}_e + f_i \vec{\mu}_i)/2 = - n \wt{\mu} \hhat$.  The collisionality factor $f$, defined using \beq \label{eqn:fs}
f_s = \dfrac{\omega_s}{\nu_s} \left(1 - e^{- \nu_s / \omega_s} \right) \left \lbrace \begin{array}{lc} \rightarrow 1 & \mathrm{for}\; \omega_s \gg \nu_s \;, \\ \rightarrow \omega_s / \nu_s & \mathrm{for}\; \omega_s \ll \nu_s \;, \end{array} \right.
\eeq where $\nu_s \equiv \nu_{s s} + \nu_{s k}$, represents collisional disruption of the gyromotion and is normalized so that $f p = \sum_s f_s p_s$.  Using $\ptild \equiv p / \mu_0$, the general magnetization model \beq
M = \dfrac{f_e \ptild_e}{H-M_i - \alpha_0 M_e} + \dfrac{f_i \ptild_i}{H-M_e- \alpha_0 M_i} = \dfrac{f \ptild}{H - \alpha M} 
\eeq has solution $M/H = (1 - \sqrt{1 - 4 \alpha f \ptild / H^2})/2 \alpha$ for $0 < \alpha \leq 1$ and $\beta$-limit $\beta \equiv 2 \ptild / H^2 \leq 1/ 2 \alpha f$ on the ratio of the kinetic to the magnetic pressure, which goes to $1/2$ for $\alpha, f \rightarrow 1$ and to $1/f$ for $\alpha \rightarrow 1/2$.  As $\omega_s$ is in terms of $\mbf{B}_s$, an iterative approach to the collisionality factor may be defined by $M_n (H-\alpha M_n) = f_n \ptild$, starting at $f_0 = 1$ with $f_n = f_n(M_{n-1})$, and for a dense, magnetically confined plasma we find $f$ remains very close to unity.

The sum of Equations~(\ref{eqn:fluid2}) gives the equilibrium net force balance equation \beq \label{eqn:equileqn}
\mbf{C}_+ + \del p_+ = \mbf{J} \times \mbf{B} + \mu_0 \del (\mbf{M}_+ \cdot \mbf{H}) \;,
\eeq now including $\mbf{F}_M$, and their difference the generalized Ohm's law equation \beq \label{eqn:ohmseqn}
\mbf{C}_- + \del p_- = n_0 e \left( \mbf{V}_i + \mbf{V}_e \right) \times \mbf{B} + \mu_0 \del (\mbf{M}_- \cdot \mbf{H}) + 2 \left[ n_0 e \mbf{E} - \mbf{F}_{e i} \right]\;,
\eeq where $n_0 e (\mbf{V}_i+\mbf{V}_e) = 2 n_0 e \mbf{V} - n_0 (m_i-m_e) \mbf{J}/\rho_m $ and $\{\mbf{V}_f,\mbf{J}_f\}$ are replaced by $\{\mbf{V},\mbf{J}\}$ everywhere except in the friction term, $p_\pm \equiv p_i \pm p_e$ and $\mbf{M}_\pm \equiv f_i \mbf{M}_i \pm f_e \mbf{M}_e$, and the convective terms $\mbf{C}_{+,-} \equiv n_0 \left[ m_i \left( \mbf{V}_i \cdot \del \right) \mbf{V}_i \pm m_e \left( \mbf{V}_e \cdot \del \right) \mbf{V}_e \right]$ are given by \bea \label{eqn:convterms}
\mbf{C}_+ =& n_0 \left(m_i + m_e\right) & \left[ \left( \mbf{V} \cdot \del \right) \mbf{V} + \dfrac{m_e m_i}{e^2} \left(\dfrac{\mbf{J}}{\rho_m} \cdot \del \right) \dfrac{\mbf{J}}{\rho_m} \right] ,\\
\mbf{C}_- =& n_0 \left(m_i - m_e\right) & \left[ \left( \mbf{V} \cdot \del \right) \mbf{V} - \dfrac{m_e m_i}{e^2} \left(\dfrac{\mbf{J}}{\rho_m} \cdot \del \right) \dfrac{\mbf{J}}{\rho_m} \right] \\ & + \dfrac{2 n_0 m_e m_i}{e} & \left[ \left( \mbf{V} \cdot \del \right) \dfrac{\mbf{J}}{\rho_m} + \left(\dfrac{\mbf{J}}{\rho_m} \cdot \del \right) \mbf{V} \right] \;.
\eea  Specializing consideration to an infinite, axisymmetric plasma column in $(r,\theta,z)$ coordinates with $\del \rightarrow (\partial/\partial r,0,0)$ and applied fields $\mbf{E}_0 = E_0 \zhat$ and $\mbf{B}_0 = B_0 \zhat$, we define profiles for the density and temperature with profile parameter $a_X$ as $n_0(r) = n_0(0) + [n_0(r_1)-n_0(0)](r/r_1)^{a_{n_0}}$ for $r_1$ at the boundary of consideration and similarly for $\Tbar_{e,i}$.  Equilibrium requires divergence-free momentum and current, $\divr (\rho_m \mbf{V}) = \divr \mbf{J} = 0$, and the convective terms reduce to $(\mbf{A} \cdot \del) \mbf{B} \rightarrow - A_\theta B_\theta \mbf{r}/r^2$.  The $\thetahat$ and $\zhat$ components of Equation~(\ref{eqn:ohmseqn}) give the free current in terms of the applied electric field $\mbf{J}_f = (0,0,n_0 e^2 E_0 / m_e \nu_{ei})$, from which the enclosed current $I_z(r) = \int_0^r dr' 2 \pi r' J_{f z}(r')$ determines the free field $H_f(r) = I_z(r)/2 \pi r$ such that $\mbf{H} = \mbf{H}_0 + H_f \thetahat$.  Our logic follows $\mbf{E}_0 \rightarrow \mbf{J}_f \rightarrow \mbf{H}_f + \mbf{H}_0 \rightarrow \mbf{M} \rightarrow \mbf{J}_d$, where the diamagnetic current is given by \beq \label{eqn:Jd}
\mbf{J}_d \equiv \curl \mbf{M} = - \curl \left(\MoH \mbf{H}\right) = \mbf{H} \times \del \MoH - \MoH \curl \mbf{H} \;,
\eeq and similarly for $\mbf{V}_d$.  Scattering times $\tau_{s k} \equiv 1/ \nu_{s k}$ are calculated using common formulas~\cite{tokamaks-2004,physics-nrl}.  The mass flow velocity $\mbf{V}$ is then found from the radial components of Equations~(\ref{eqn:equileqn}) and (\ref{eqn:ohmseqn}).  Results for a typical analysis are found in Figures~(1) and (2) with the numerical parameters given in the first figure caption.  Note that the theory above accounts naturally for the ``spontaneous'' generation of a flow velocity as a consequence of the equilibrium force balance equations and offers an alternative to the neoclassical approach for the analysis of plasma experiments~\cite{frc-pop-2006,solomonetal-pop-2006}.

In conclusion, we have found that Gauss's law may not be extracted from the covariant field tensor equation without introducing serious mathematical inconsistencies.  The expression of Maxwell's theory in intrinsic, geometric form holds at the classical level, and suitably quantized its predictions have been verified to an accuracy of parts per billion.  To conjecture a range of energy density at which it fails to hold is inconsistent with experience of both the everyday world and precision experimental tests.  Reinstating Gauss's law lets one predict both the momentum and current within a plasma from the principles of electromagnetic hydrodynamic theory, allowing for the experimental testing of the hypothesis that Maxwell's equations really do apply to the physics of many-body systems.





\newpage
\begin{figure}
\includegraphics[width=3.375in]{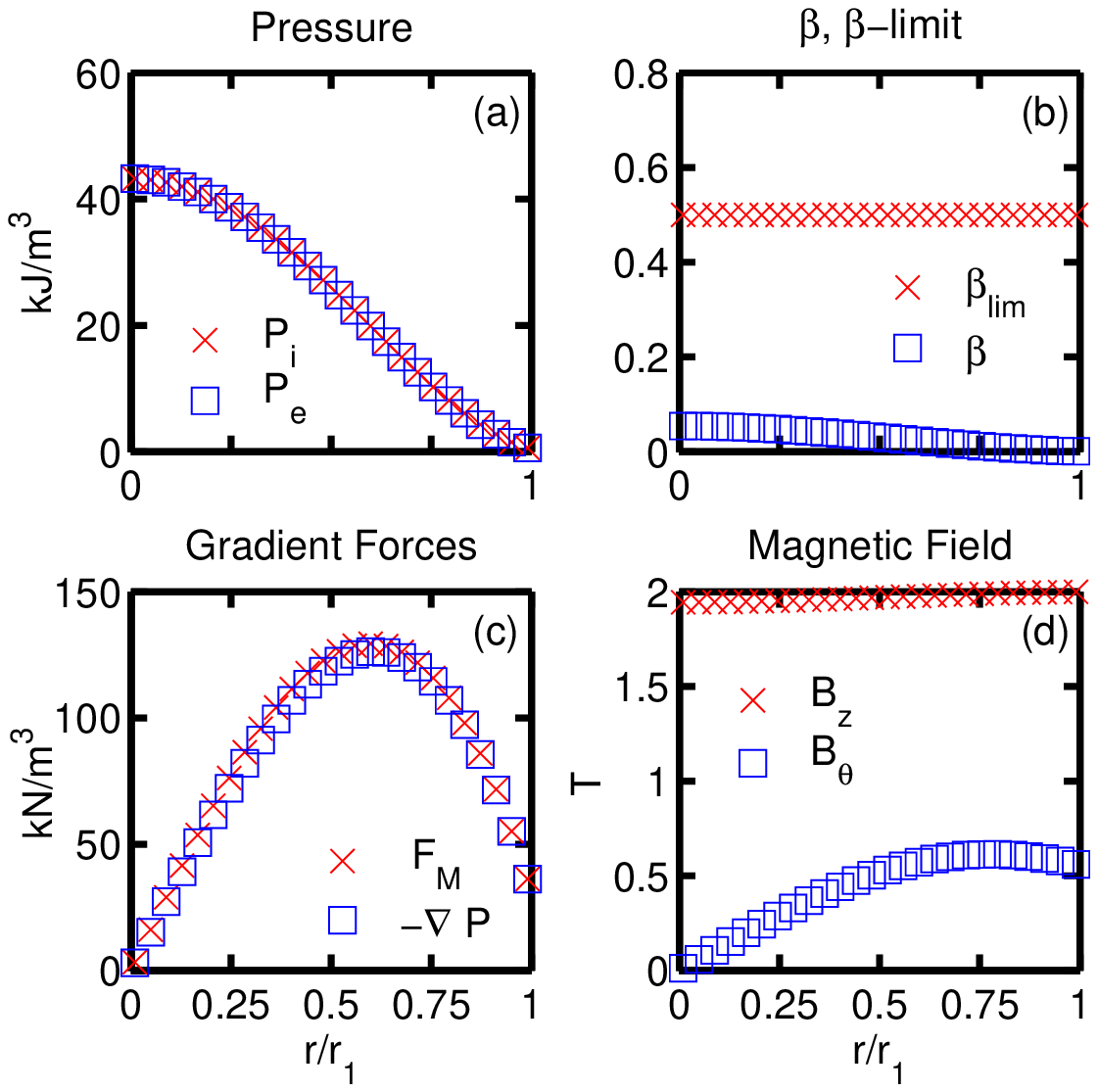}
\caption{\label{fig1} (Color online.)  Results from a typical analysis of magnetically confined, fully ionized deuterium with $r_1 = 1$m, applied electric field $E_0 = 20$mV/m, and applied magnetic field $B_0 = 2$T.  (a) Input pressure profiles with central density $n_0(0) = 9\times10^{19}/{\mathrm m}^3$, central temperature $\Tbar_{e,i}(0) = 3$keV, profile parameter $a_X = 2$, and pedestal parameter $X(0)/X(1) = 10$.  (b) Calculated $\beta$-limit and $\beta$ with a central value of 5.4\%.  (c) The magnetization force equals or exceeds the pressure gradient force in the equilibrium equation.  (d) The net magnetic field is reduced from the free current value by the magnetization.}
\end{figure}

\begin{figure}
\includegraphics[width=3.375in]{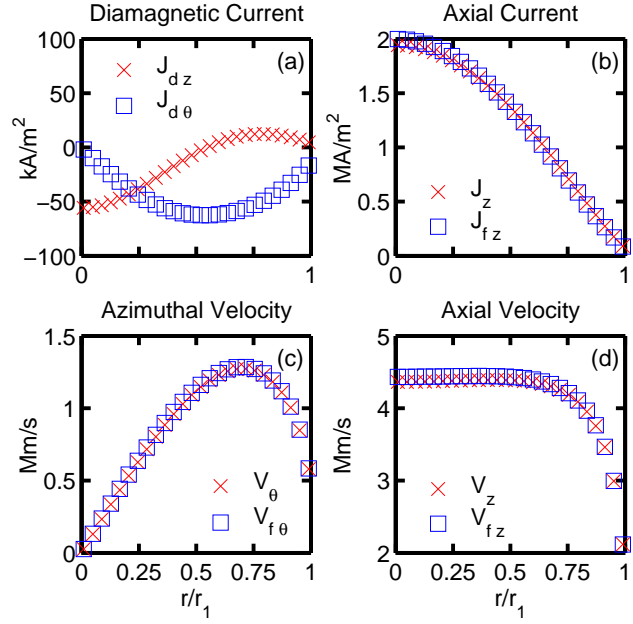}
\caption{\label{fig2} (Color online.)  Results from a typical analysis of magnetically confined, fully ionized deuterium with $r_1 = 1$m and total enclosed free current $I_z = 2.8$MA.  (a) The diamagnetic current has both an azimuthal and axial component.  (b) The net axial current is reduced from the free current value by the axial diamagnetic current.  (c) \& (d) The effect of the gyromotion on the plasma flow velocity is small but non-negligible.  Here, $\mbf{V}_f$ is the flow with neglect of the gyromotion and $\mbf{V}$ includes the drift velocity $\mbf{V}_d$, found to be on the order of mm/s.}
\end{figure}

\end{document}